\documentclass[10pt, preprint]{emulateapj}

\usepackage{color}

\usepackage{style_wei_symbols}
\usepackage{style_wei}

\begin{document}

\title{An Intriguing Chromospheric Jet Observed by {\it Hinode}: I. Fine Structure Kinematics 
and Evidence of Unwinding Twists}

\author{Wei Liu\altaffilmark{1}\altaffilmark{2}, Thomas E.~Berger\altaffilmark{1}, Alan M.~Title\altaffilmark{1},
and Theodore D.~Tarbell\altaffilmark{1}}


\altaffiltext{1}{Lockheed Martin Solar and Astrophysics Laboratory, Department ADBS, 
Building 252, 3251 Hanover Street, Palo Alto, CA 94304}
\altaffiltext{2}{W.~W.~Hansen Experimental Physics Laboratory, Stanford University, Stanford, CA 94305}

\shorttitle{Chromospheric Jet Observed by {\it Hinode}}
\shortauthors{Liu et al.}

\journalinfo{Accepted by ApJ Letters 2009}
\submitted{Received 2009 September 05; accepted 2009 October 27}

\begin{abstract}	

We report a chromospheric jet lasting for more than 1~hr observed by {\it Hinode} Solar Optical Telescope 
in unprecedented detail.
The ejection occurred in three episodes separated by 12--14~min, with the amount and velocity of material decreasing with time.
  The upward velocities range from 438 to $33 \kmps$, while the downward velocities of the material falling back
have smaller values (mean: $-56 \kmps$) and a narrower distribution (standard deviation: $14 \kmps$).
  The average acceleration inferred from parabolic space-time tracks is $141 \m \, {\rm s^{-2}}$, a fraction of the solar gravitational
acceleration. 
The jet consists of fine threads ($0\farcs5$--$2\arcsec$ wide), which exhibit coherent, oscillatory transverse motions 
perpendicular to the jet axis and about a common equilibrium position. 
  These motions propagate upward along the jet, with the maximum phase speed of $744 \pm 11 \kmps$
at the leading front of the jet.
  The transverse oscillation velocities range from $151$ to $26 \kmps$, amplitudes from $6.0$ to $1.9 \Mm$, and periods from $250$ to $536 \s$.
  The oscillations slow down with time and cease when the material starts to fall back. 
The falling material travels along almost straight lines in the original direction of ascent,
showing no transverse motions.
These observations are consistent with the scenario that the jet involves
untwisting helical threads, which rotate about the axis of a single large cylinder
and shed magnetic helicity into the upper atmosphere.

\end{abstract}


\section{Introduction}
\label{sect_intro}

Transient, small-scale ejections of plasma from the lower atmosphere are common manifestations
of solar activity. Cool plasma ejections were observed as emission in \Ha 
or absorption at other wavelengths and historically called {\it surges} \citep{NewtonH.surge-discovery.1934MNRAS..94..472N}.
Hot ejections were usually called {\it jets} and observed as emission in
ultraviolet \citep{Brueckner.Bartoe.UVjet1983ApJ...272..329B}, 
extreme ultraviolet \citep[EUV;][]{Alexander.Fletcher.jet.1999SoPh..190..167A}, 
and soft X-rays \citep{ShibataK.1st-SXT-jet.1992PASJ...44L.173S, StrongK.SXT-jet1992PASJ...44L.161S}.
Some observations \citep{SchmiederB.cool-surge.hot-jet1988A&A...201..327S,
ChaeQiu.jet.surge1999ApJ...513L..75C, JiangYC.surge.jet.twist2007A&A...469..331J} have indicated that cool surges and hot jets 
are closely related in space and time, but their physical relationship has not been established.	
Torsional motions or helical features have long been seen in surges or jets
\citep[e.g.,][]{XuAA.surge-rotate.1984AcASn..25..119X, Kurokawa.untwist-filamt1987SoPh..108..251K,
ShimojoM.jet-stat.1996PASJ...48..123S, Patsourakos.EUVI-jet2008ApJ...680L..73P},
while their exact causes remain unclear.
From now on, we refer to both surges and jets with a general term {\it jets}, unless otherwise noted.

Solar jets are commonly associated with		
flux emergence \citep{Roy.surge.magnetic.1973SoPh...28...95R}	
or moving magnetic features \citep{Gaizauskas.movingBfield.surge1982AdSpR...2...11G, BrooksDH.surge2007ApJ...656.1197B}.
Models involving magnetic reconnection have thus been proposed 	
\citep[e.g.,][]{RustD.surge-reconn.1968IAUS...35...77R, YokoyamShibata.jetModel1995Natur.375...42Y}.
Mechanisms \citep[see review in][]{Canfield.surge-jet1996ApJ...464.1016C}		
suggested for accelerating jet material to 10--1000~$\kmps$
include reconnection outflows driven by the slingshot effect of magnetic tension
\citep{YokoyamShibata.jetModel1995Natur.375...42Y}, 
the pressure gradient behind the shock formed by reconnection outflows
\citep{YokoyamShibata.jetModel1996PASJ...48..353Y, Tarbell.Ryutova.jet.1999ApJ...514L..47T},
chromospheric evaporation caused by heating from the associated flare \citep{ShimojoM.jet-phys.2000ApJ...542.1100S},
and relaxation of magnetic twists \citep{Shibata.Uchida.helic-jetMHD.1985PASJ...37...31S}.

\hinode \citep{Kosugi.Hinode2007SoPh..243....3K}, with its superior resolutions,
has provided unprecedented details and spurred renewed interest in solar jets. They were 
found to be ubiquitous on various spatial and temporal scales in
the \ion{Ca}{2} H line \citep{Shibata.CaHjet.2007Sci...318.1591S, Nishizuka.giantCaHjet.2008ApJ...683L..83N},
EUV \citep{Culhane.EIS-jet2007PASJ...59S.751C, Moreno-Insertis.EISjet2008ApJ...673L.211M},
and soft X-rays \citep{Savcheva.XRT.jet.stat2007PASJ...59S.771S, NittaN.jetHe3.2008ApJ...675L.125N}. 
%
In this Letter, we report an intriguing jet observed by \hinode in great detail, 
and attempt to obtain new clues to some unanswered questions mentioned above.
 \begin{figure*}[thb]      
 \epsscale{1.2}	
 \plotone{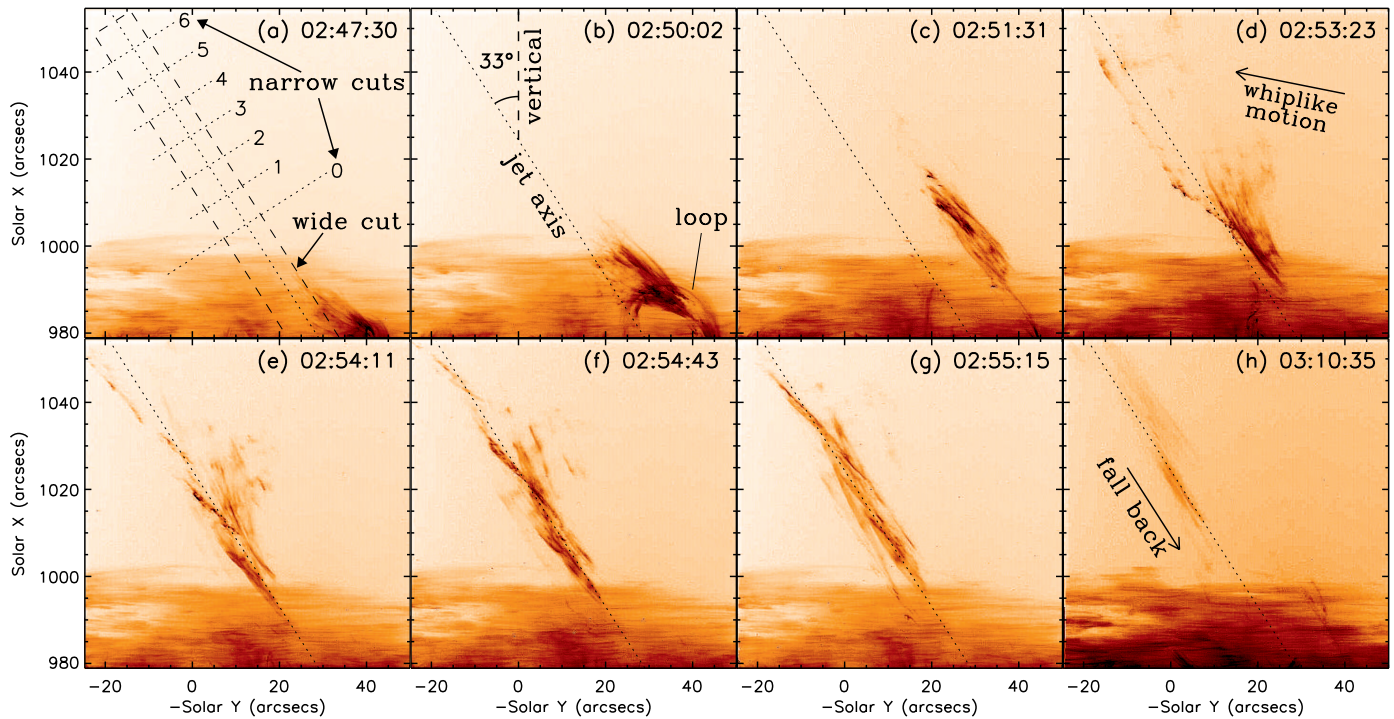}
 \caption[]{
 Negative \hinode \ion{Ca}{2} H images.		
 The diagonal dotted line 	
 marks the jet axis. The dashed box indicates a $10\arcsec$ wide cut along the axis for the time-distance
 diagram shown in Figure~\ref{para.eps}(a),		
 while the numbered dotted lines represent $1\arcsec$ narrow cuts 
 perpendicular to the axis (see online movie~1).
 } \label{maps.eps}
 \end{figure*}
%

\section{Observations and Data Analysis}
\label{sect_obs}

The chromospheric jet under study occurred from 02:40 to 04:20~UT on 2007 February 9 in AR~10940 on the west limb,
accompanied by a \goes A4.9 flare.		
It was observed by the Solar Optical Telescope \citep[SOT;][]{Tsuneta.SOT.2008SoPh..249..167T}		
and X-Ray Telescope 	
on board \hinodeA, and by \traceA, \stereoA, \hsiA, and several ground-based observatories.
The SOT observations were obtained in the \ion{Ca}{2} H line passband,
sensitive to chromospheric temperatures of 1--2$\E{4}\K$. 	
The high spatial resolution ($\sim$$0\farcs 2$) 	
and cadence (8~s) allowed us to investigate the fine structures and kinematics of the jet, 
which are the focus of this Letter. A comprehensive study of this event using multiwavelength data
will be presented in the future.	

We processed Ca images with the standard \texttt{fg\_prep} routine,	
and cross-correlated neighboring images to correct for pointing drifts, 
with a sub-arcsecond accuracy over a 2~hr duration. The absolute solar coordinates were 
determined by fitting the limb and are thus approximate with uncertainties in the roll angle.
We then applied a radial filter to reduce the contrast between the bright disk and faint emission 
above the limb, and finally deconvolved the images with the SOT point spread function determined using the lunar limb 
during eclipses (C.~DeForest, 2009, private communication) to sharpen fine features. 

A sample of the resulting images are rotated clockwise by $90\degree$ and
plotted in Figure~\ref{maps.eps}, showing the evolution of the jet.
We find that the precursor of the jet appeared at 02:40~UT as a bundle of fine material		
threads (typical widths: $0\farcs5$--$2\arcsec$)		
at $\sim$40$\degree$ from the photosphere extending to a height of $\sim$20$\arcsec$. 	
With these threads exhibiting oscillatory transverse motions across its axis,	
this bundle rose up, first gradually and then rapidly, and an overarching loop grew
simultaneously (Figure~\ref{maps.eps}(b)). 
At 02:51:31~UT, the bundle started to swing back to the left in a whiplike manner (Figure~\ref{maps.eps}(d)),
similar to that in simulated cool ($10^4\K$) jets \citep{YokoyamShibata.jetModel1995Natur.375...42Y}.
Meanwhile, with their 	
transverse motions continuing, the threads were ejected upwards (Figures~\ref{maps.eps}(d)--(g)).
Later in the event up to 04:20~UT, material bound by gravity fell back to the chromosphere along the
original paths of ascent.		
In contrast to those of the ascending material, the streamlines of the falling material were almost straight lines, 
with no detectable transverse motions (Figure~\ref{maps.eps}(h)).		

To quantify the kinematics of the jet, we constructed an orthogonal coordinate system based on the
jet orientation. We first visually determined the axis passing through the center of
the jet when it was fully developed and collimated (e.g., Figure~\ref{maps.eps}(g)). 
This axis (dotted diagonal line in Figure~\ref{maps.eps}) is located in the sky plane and at
$33 \degree$ from the local vertical on the equator, and its intersection with the photosphere
is defined as the origin $O$ for measuring the distance $s_\parallel$ along
the jet. We used a $10\arcsec$ {\it wide} cut (dashed box in Figure~\ref{maps.eps}(a)) {\it along} the 
axis, and averaged pixels across the width to obtain the time-distance diagram 
for the axial motions. Likewise, we selected 7 {\it narrow} cuts ($1\arcsec$ wide, numbered 0--6) {\it perpendicular}
to and centered at the jet axis to study the transverse motions. 
These narrow cuts are uniformly spaced by $8\arcsec$ 		
and Cut~0 is positioned at $s_\parallel=39 \arcsec$ 	
from $O$. The distance $s_\perp$ across the jet along the narrow cuts is measured from upper-right to lower-left
in Figure~\ref{maps.eps}(a) with $s_\perp=0$ located at the jet axis.

\subsection{Axial Motions along Jet}
\label{subsect_para}

\begin{figure}[thbp]      
 \epsscale{1.2}	
 \plotone{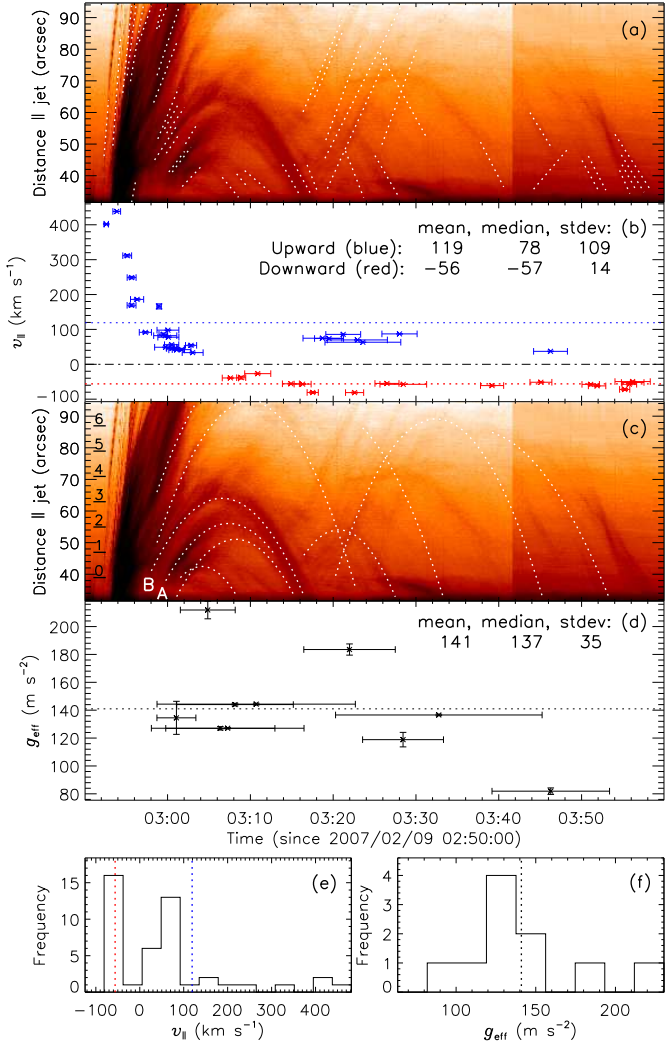}
 \caption[]{ Kinematics along the jet.
  (a) Time-distance diagram of Ca intensity from the wide cut	
 shown in Figure~\ref{maps.eps}(a), overlaid with linear fits to individual tracks (see online movie~2).
 The increase in the background level at 03:41:40~UT is due to the pointing jump toward the disk center
 that possibly resulted in more scattered light in the telescope.
  (b) Velocities of the above fits vs.~time. The horizontal bars indicate
 the time intervals of the fits. The horizontal dotted lines mark the
 means of the upward ($>$0) and downward ($<$0) velocities.
  (c) Same as (a) but overlaid with parabolic fits to selected tracks.	
 The short bar underneath each number (0--6) on the left marks the position of the corresponding
 narrow cut shown in Figure~\ref{maps.eps}(a).
  (d) Same as (b) but for the effective gravitational acceleration $g_{\rm eff}$ inferred
 from the fits in (c).
  (e) and (f) Histograms of the quantities		
 shown in (b) and (d), with the means marked by the same dotted lines.
 } \label{para.eps}
 \end{figure}

The axial motions of material along the jet are represented by the time-distance diagram 
(Figure~\ref{para.eps}(a)) from the wide cut.
The upward (downward) tracks correspond to ascending (falling) material. 
Some tracks show deceleration and change from upward to downward within the SOT field of view (FOV).
We visually identified all unambiguous, straight-shaped portions of upward tracks just before deceleration and 
of downward tracks at the lowest altitudes where they are visible. We fitted these tracks linearly 	
and show the fits and velocities in Figures~\ref{para.eps}(a) and (b).
%

As is evident, there are three episodes of material ejections, separated by 12--14~min.
The number of tracks and amount of material of these episodes decrease with time;
so do the mean and scatter of the upward velocities. 
During the first episode alone, the velocity rapidly decreases 	
from $438.3 \pm 0.4$ to $33 \pm 1 \kmps$ within 9~min.
The mean and standard deviation of all the upward velocities are $119$ and $109 \kmps$, 
respectively. In contrast, for the falling material, the downward velocities have less distinct
episodes with a much smaller mean and scatter, and these values are $-56$ and $14 \kmps$ (Figures~\ref{para.eps}(b) and (e)).

For those tracks showing clear concave curvatures (see Figure~\ref{para.eps}(c)),
we fitted their positions with a parabolic function of time $t$, 
 \beq
  s_\parallel(t) = s_{\parallel 0} + v_{\parallel 0} t - {g_{\rm eff} \over 2} t^2 \,,
 \label{parabola_eq} \eeq
where $g_{\rm eff}$ is the
effective gravitational acceleration, and $s_{\parallel 0}$ and $v_{\parallel 0}$ are the initial position
and velocity at $t=0$. The fitted values of $g_{\rm eff}$ (Figures~\ref{para.eps}(d) and (f)) have a
mean and standard deviation of $141$ and $35 \mpss$, respectively.
%
The component of the solar gravitational acceleration along the jet axis 	
ranges from $231 \mpss$ at the photosphere to $197 \mpss$ at the top of the FOV
(radial position $r=1054\arcsec$). 
Our fitted $g_{\rm eff}$ is only a fraction of these values.	
Possible explanations for this apparent discrepancy include that:	
 (1) the jet is likely to be oriented out of the sky plane
due to the line-of-sight effect, such that the discrepancy 	
would be smaller,
and (2) the force driving the mass ejection, possibly at lower levels late in the event, 
could still reduce the effect of gravity.
Note that \citet{Roy.surge.dynamics.1973SoPh...32..139R} also found in \Ha surges 	
the acceleration of falling material being less than free-fall, but 
the deceleration of ascending material being greater than gravity alone.

\subsection{Transverse Motions across Jet}
\label{subsect_perp}

 \begin{figure}[thbp]      
 \epsscale{1.2}	
 \plotone{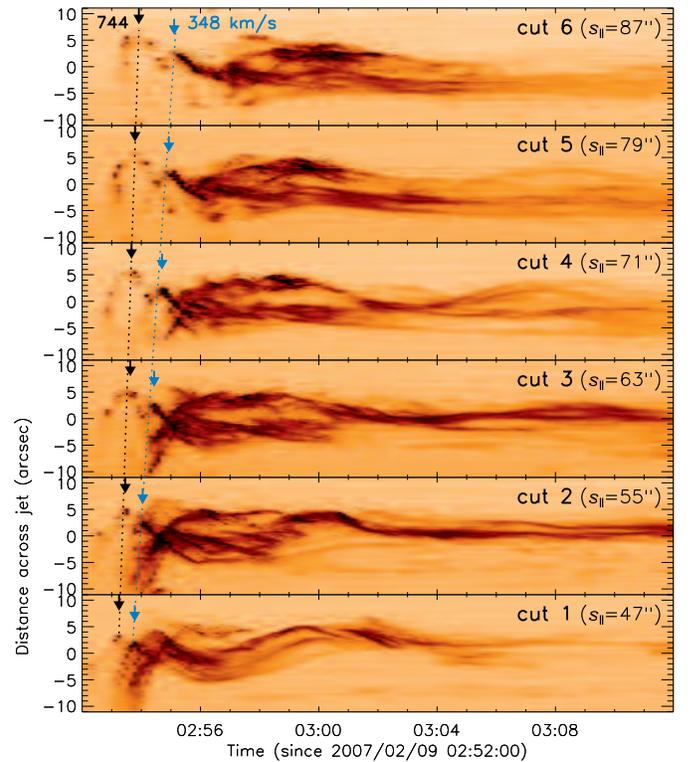}
 \caption[]{Time-distance diagram from $1\arcsec$ narrow cuts 1--6 perpendicular to the jet shown in Figure~\ref{maps.eps}(a).
 The two arrows in each panel point to the crests	
 of the first two inverted-V shaped tracks, indicating delays toward higher altitudes. The dotted lines are linear fits 
 to the delays; the resulting phase speeds are labeled.
 } \label{perp.eps}
 \end{figure}
%
The time-distance plot (Figure~\ref{perp.eps}) from the narrow cuts clearly
shows oscillatory transverse motions perpendicular to the jet during the first episode of material ejection. 
These oscillations propagate along the jet to higher altitudes, as indicated by the time delays across different cuts.
We identified the first inverted-V shaped  
track in each panel, 
and by linearly fitting the axial positions $s_\parallel$ of the cuts 	
vs.~the occurrence times of the track crest (indicated by the dark arrows), 
we obtained a phase speed of $v_{\rm ph}=744 \pm 11 \kmps$. 
The propagation of the second such track (indicated by the blue arrows) is
50\% slower at $v_{\rm ph}=348 \pm 5 \kmps$. 
We also note that the oscillation amplitude decreases with time in each cut. 
At a higher cut, the separations between the threads become larger
and various phase delays appear in their oscillations.

\begin{figure}[thbp]      
 \epsscale{1.2}	
 \plotone{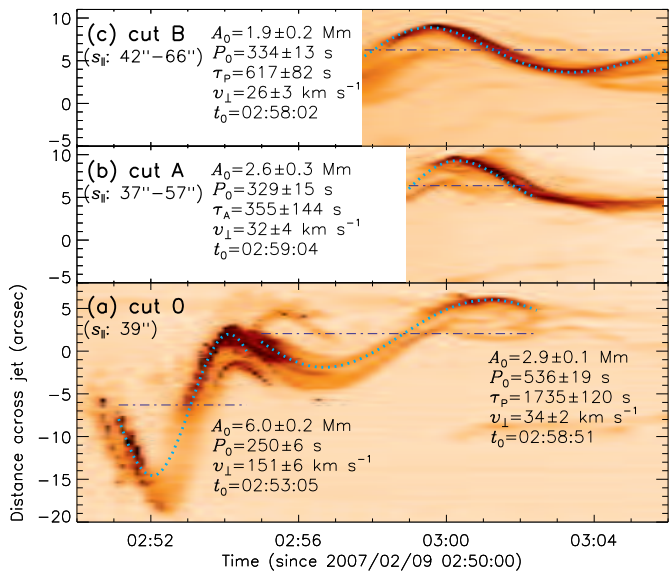}
 \caption[]{	
 Same as Figure~\ref{perp.eps} but for
 (a) Cut~0 shown in Figure~\ref{maps.eps}(a),
 (b) and (c) cuts co-moving with the ascending material on the upward tracks~A and B
 shown in Figure~\ref{para.eps}(c). 
 The ranges of the cut position $s_\parallel$ along the jet axis are labeled.
 The dotted lines are fits to fine tracks	
 with a damped sine function (see text), whose parameters
 are shown in the legends (with $\tau_A$ or $\tau_P=\infty$ omitted). 
 The dot-dashed lines mark the fitted equilibrium positions.	
 } \label{perp-fit.eps}
 \end{figure}

To quantify the transverse oscillations, we used the time-distance plot of Cut~0 as an example
(Figure~\ref{perp-fit.eps}(a)), in which such motions are the most pronounced. We fitted the
piecewise oscillatory tracks with a damped sine function:
 \beqa 
 s_\perp(t) & = &  s_{\perp 0} + A(t) \sin \left[ 2 \pi \int_{t_0}^t {d\xi \over P(\xi)} \right]  \nonumber      \\
   & = & s_{\perp 0} + A(t) \sin \left[ 2 \pi \tau_P \left( {1 \over P_0 } - {1 \over P(t)} \right) \right]  \,,
 \label{sin_eq} \eeqa
where $A(t) = A_0 \exp[-(t-t_0)/\tau_A]$ is the amplitude,
and $P(t) = P_0 \exp[(t-t_0)/\tau_P]$ is the period, 
with $\tau_A$ and $\tau_P$ being their damping time.
We find a rapid oscillation at the beginning of the event, corresponding to the
rise and whiplike swing of the entire bundle mentioned earlier. The fitted transverse
velocity at the equilibrium position near the middle of this track duration is $v_\perp=151 \pm 6 \kmps$,
and the amplitude and period at $t=t_0$ are $A_0= 6.0 \pm 0.2 \Mm$ and $P_0= 250 \pm 6 \s$.
This is followed by a slower oscillation with the amplitude halved and the period doubled,
giving a velocity of $v_\perp=34 \pm 2 \kmps$.		

We also moved the $1\arcsec$ narrow cut with time along the jet axis to keep up with the
upward moving material and to remove the effect of its axial motion.
The time-distance diagrams of such co-moving Cuts A and B
are shown in Figures~\ref{perp-fit.eps}(b) and (c) for the upward 
tracks~A and B marked in Figure~\ref{para.eps}(c), respectively.
Their damped sine fits yield transverse velocities similar to
that of the second oscillation of the stationary Cut~0 noted above.
%


\section{Discussions}
\label{sect_discuss}

The simplest and most intuitive scenario that can tie together the above observations 
is that the jet is composed of helical threads undergoing untwisting spins, 
as proposed by \citet{Shibata.Uchida.helic-jetMHD.1985PASJ...37...31S, Shibata.Uchida.helic-jet.1986SoPh..103..299S}
and \citet{Canfield.surge-jet1996ApJ...464.1016C}.
In this scenario, a highly-twisted emerging flux rope or existing filament
reconnects with less twisted ``open" field lines, along which 	
the stored twists and magnetic helicity are transported to the upper atmosphere. 
We interpret our observations in this framework as follows.		

{\bf Chirality.} The alternate and mutual shadowing of multiple threads consistently over time suggests that 
the threads oriented as crossing the jet axis from lower right to upper left are on the
front side (closer to the observer), while those crossing from lower left to upper right
are on the far side (Figures~\ref{maps.eps}(d)--(g)). 	
These threads, assumed to represent magnetic field lines,
are thus {\it left-handed} screws. The former threads are seen to traverse the jet axis from left to right, 
while the latter from right to left. Their transverse motions (see online movie~1)
are thus counter-clockwise rotations when looking from above along the jet axis.
The trajectories of the material, considering its upward axial motion, are thus
{\it right-handed} helices. This is consistent with the unwinding of twisted
field lines of the opposite handedness inferred here, as indicated by surge observations
\citep{Canfield.surge-jet1996ApJ...464.1016C, Jibben.Canfield.twist-surge-stat2004ApJ...610.1129J}
and 3D MHD simulations \citep{PariatE.twist-jet-MHD.2009ApJ...691...61P}.

{\bf Oscillatory transverse motions}%
 \footnote{The apparent transverse motions at cuts~1--6 shown in Figure~\ref{perp.eps} 
 result from the rotation about and the translation along the jet axis of
 individual unwinding helical threads. However, at the lowest narrow cut (Cut 0), since the bundle of threads
 is packed as a whole 	
 and extends almost parallel to the jet axis, 
 Figure~\ref{perp-fit.eps}(a) represents mainly the true rotation of the entire bundle. 
 So do Figures~\ref{perp-fit.eps}(b) and (c) for the co-moving cuts.}
across the jet, in general, can be understood as the projections of rotations about an axis in the sky plane.
The initial rotation of the entire bundle, in particular, appears in a counter-clockwise coning manner,
and its velocity $v_\perp=151 \pm 6 \kmps$ is 5-fold greater than 
the other rotations during this event, but comparable to the $120 \kmps$ velocity
of whiplike or transverse motions reported for surges or jets \citep{Canfield.surge-jet1996ApJ...464.1016C, ShimojoM.XRT-fine-jet.2007PASJ...59S.745S}.
The rest of the rotational velocities around $30 \kmps$ and periods in the 330--540~s range
are comparable to those of transverse oscillations found in 
X-ray jets \citep{Cirtain.XRTjet2007Sci...318.1580C},
prominences \citep{Okamoto.Alfven-wave.prominence.2007Sci...318.1577O},
and coronal loops \citep{Ofman.Wang.SOT-wave.2008A&A...482L...9O}, 
which were interpreted as signatures of transverse MHD or Alfv\'{e}n waves.
The angular velocities inferred from Doppler measurements in \Ha surges 
are on the order of $10^{-3} \, {\rm rad} \, \ps$
\citep{XuAA.surge-rotate.1984AcASn..25..119X, GuXM.helical-surge1994A&A...282..240G,
Jibben.Canfield.twist-surge-stat2004ApJ...610.1129J},		
which translates to a period of $6\E{3} \s$.
These rotations are 10 times	
slower than reported here, although their axial velocities are only fractionally smaller.	

{\bf Axial motions.}
Propagating torsional Alfv\'{e}n waves associated with unwinding twists
can explain the observed delays of the transverse motions with distance along the jet.
The initial phase speed $v_{\rm ph}=744 \pm 11 \kmps$ (Figure~\ref{perp.eps})
would correspond to the untwisting front or progressing pinch front which, according to
\citet{Shibata.Uchida.helic-jetMHD.1985PASJ...37...31S}, propagates at the Alfv\'{e}n speed.	
Proposed driving mechanisms of jet material also include
torsional Alfv\'{e}n waves \citep{PariatE.twist-jet-MHD.2009ApJ...691...61P} and
the ${\bf J} \times {\bf B}$ force in the untwisting helices \citep{Shibata.Uchida.helic-jetMHD.1985PASJ...37...31S}.
The decrease with time of the upward velocities as shown in Figures~\ref{para.eps}(a)--(b) indicates
that the pitch spacing between adjacent helical threads increases with height. This is consistent
with the expected axial expansion of unwinding screws. 

{\bf Temporal evolution.} We find a slowdown of the transverse motions 
(see Figures~\ref{perp.eps} and \ref{perp-fit.eps}), which accompanies the	
decrease of the axial velocity (Figure~\ref{para.eps}(b)). In particular, when the rotations of Cuts~A and B
stop (Figures~\ref{perp-fit.eps}(b) and (c)) near 03:06--03:08~UT, the corresponding 	
ascending material reaches its maximum height and starts to fall back. On its return,
the material follows paths that are almost straight lines (until $\sim$$20\arcsec$ above the photosphere)
along the jet axis and shows no detectable transverse motions. Assuming the paths are along magnetic field lines, 
this means very little or no twist in the post-eruption field. 
This provides further evidence for the causal link between
the jet and unwinding twists 	
and for the role played by the jet in shedding away twists or magnetic helicity.

An alternative to our rotation interpretation is planar waves, which seem, however, difficult
to explain some of the observations (see online movie~1):
(1) Multiple threads are seen to undergo coherent, oscillatory transverse motions, one following another 
with consistent phase delays and reversing their direction of motion at roughly the same distance from the jet axis. 
This is consistent with surge observations \citep{Canfield.surge-jet1996ApJ...464.1016C}	
and can be readily understood if these threads are distributed on a single rotating cylindrical surface.
(2) Likewise, some sheet-shaped material volumes periodically change their appearances in a way 
as if they are on the surface of a rotating cylinder.		
To distinguish between these two interpretations with certainty, simultaneous high-resolution Ca images 
and \Ha Doppler measurements are needed and can both be obtained with SOT (but unavailable for this event).

\section{Conclusions}	
\label{sect_conclude}

We have presented kinematic measurements of both axial and transverse motions
of a chromospheric jet observed by \hinode SOT at high spatial and temporal resolutions.
This study complements previous low resolution counterparts of \Ha surges 	
and EUV or X-ray jets, and offers new insights to this type of phenomena. 
Our major results and interpretations are as follows.

\begin{enumerate}

\item	
The ejection occurs in three episodes separated by 12--14~min,
rather than continuously. The amount and velocity of ejected material decrease with time.	
The ejecting velocities have a wide range from 438 to $33 \kmps$,
while the velocities of material falling back 		
have a narrow range (mean: $-56 \kmps$ and standard deviation: $14 \kmps$).
The acceleration inferred from parabolic tracks in the time-distance diagram has a mean 
of $141 \m \, {\rm s^{-2}}$, a fraction of the solar gravitational acceleration.

\item	
The jet consists of $0\farcs5$--$2 \arcsec$ thick fine threads,
which exhibit oscillatory transverse motions across the jet about a common equilibrium position.
These oscillations have velocities ranging from $151 \pm 6$ to $26 \pm 3 \kmps$,
amplitudes from $6.0 \pm 0.2$  to $1.9 \pm 0.2 \Mm$,
and periods from $250 \pm 6$ to $536 \pm 19 \s$.
The upward propagation of the oscillations has a maximum phase speed of
$744 \pm 11 \kmps$ (comparable to the coronal Alfv\'{e}n speed) 
associated with the leading front of the jet. 

\item	
The transverse motions slow down with time and cease near the time when the material reaches its maximum height
and starts to fall back. The falling material travels along almost straight lines in the original direction of ascent,
showing no more signatures of transverse motions.

\item	
These observations are consistent with the scenario that the jet involves
unwinding of left-handed helical threads that rotate counter-clockwise about a common
axis. The untwisting wave front propagates upward at the Alfv\'{e}n speed.
The pitch spacing between adjacent helical threads increases with height, consistent
with the expected axial expansion of	
unwinding screws. The jet results in	
magnetic helicity being shed into the upper atmosphere.

\end{enumerate}

A more in-depth multiwavelength study of this event and comparison with theoretical models are underway and
will be published in the future.



\acknowledgments
{
Many thanks to Richard Shine, Tong-Jiang Wang, T.~J.~Okamoto, and Alexey Kruglov.
%
This work was supported by \hinode SOT contract NNM07AA01C.		
\hinode is a Japanese mission developed and launched by ISAS/JAXA, with NAOJ as
domestic partner and NASA and STFC (UK) as international partners. It is operated
by these agencies in cooperation with ESA and NSC (Norway). 
}





{\scriptsize
 \begin{thebibliography}{39}
\expandafter\ifx\csname natexlab\endcsname\relax\def\natexlab#1{#1}\fi

\bibitem[{{Alexander} \&
  {Fletcher}(1999)}]{Alexander.Fletcher.jet.1999SoPh..190..167A}
{Alexander}, D. \& {Fletcher}, L. 1999, \solphys, 190, 167

\bibitem[{{Brooks} {et~al.}(2007){Brooks}, {Kurokawa}, \&
  {Berger}}]{BrooksDH.surge2007ApJ...656.1197B}
{Brooks}, D.~H., {Kurokawa}, H., \& {Berger}, T.~E. 2007, \apj, 656, 1197

\bibitem[{{Brueckner} \&
  {Bartoe}(1983)}]{Brueckner.Bartoe.UVjet1983ApJ...272..329B}
{Brueckner}, G.~E. \& {Bartoe}, J.-D.~F. 1983, \apj, 272, 329

\bibitem[{{Canfield} {et~al.}(1996){Canfield}, {Reardon}, {Leka}, {Shibata},
  {Yokoyama}, \& {Shimojo}}]{Canfield.surge-jet1996ApJ...464.1016C}
{Canfield}, R.~C., {Reardon}, K.~P., {Leka}, K.~D., {Shibata}, K., {Yokoyama},
  T., \& {Shimojo}, M. 1996, \apj, 464, 1016

\bibitem[{{Chae} {et~al.}(1999){Chae}, {Qiu}, {Wang}, \&
  {Goode}}]{ChaeQiu.jet.surge1999ApJ...513L..75C}
{Chae}, J., {Qiu}, J., {Wang}, H., \& {Goode}, P.~R. 1999, \apjl, 513, L75

\bibitem[{{Cirtain} {et~al.}(2007){Cirtain}, {Golub}, {Lundquist}, {van
  Ballegooijen}, {Savcheva}, {Shimojo}, {DeLuca}, {Tsuneta}, {Sakao}, {Reeves},
  {Weber}, {Kano}, {Narukage}, \&
  {Shibasaki}}]{Cirtain.XRTjet2007Sci...318.1580C}
{Cirtain}, J.~W., et al.
  2007, Science, 318, 1580

\bibitem[{{Culhane} {et~al.}(2007){Culhane}, {Harra}, {Baker}, {van
  Driel-Gesztelyi}, {Sun}, {Doschek}, {Brooks}, {Lundquist}, {Kamio}, {Young},
  \& {Hansteen}}]{Culhane.EIS-jet2007PASJ...59S.751C}
{Culhane}, L., et al.
  2007, \pasj, 59, 751

\bibitem[{{Gaizauskas}(1982)}]{Gaizauskas.movingBfield.surge1982AdSpR...2...11%
G}
{Gaizauskas}, V. 1982, Advances in Space Research, 2, 11

\bibitem[{{Gu} {et~al.}(1994){Gu}, {Lin}, {Li}, {Xuan}, {Luan}, \&
  {Li}}]{GuXM.helical-surge1994A&A...282..240G}
{Gu}, X.~M., {Lin}, J., {Li}, K.~J., {Xuan}, J.~Y., {Luan}, T., \& {Li}, Z.~K.
  1994, \aap, 282, 240

\bibitem[{{Jiang} {et~al.}(2007){Jiang}, {Chen}, {Li}, {Shen}, \&
  {Yang}}]{JiangYC.surge.jet.twist2007A&A...469..331J}
{Jiang}, Y.~C., {Chen}, H.~D., {Li}, K.~J., {Shen}, Y.~D., \& {Yang}, L.~H.
  2007, \aap, 469, 331

\bibitem[{{Jibben} \&
  {Canfield}(2004)}]{Jibben.Canfield.twist-surge-stat2004ApJ...610.1129J}
{Jibben}, P. \& {Canfield}, R.~C. 2004, \apj, 610, 1129

\bibitem[{{Kosugi} {et~al.}(2007){Kosugi}, {Matsuzaki}, {Sakao}, {Shimizu},
  {Sone}, {Tachikawa}, {Hashimoto}, {Minesugi}, {Ohnishi}, {Yamada}, {Tsuneta},
  {Hara}, {Ichimoto}, {Suematsu}, {Shimojo}, {Watanabe}, {Shimada}, {Davis},
  {Hill}, {Owens}, {Title}, {Culhane}, {Harra}, {Doschek}, \&
  {Golub}}]{Kosugi.Hinode2007SoPh..243....3K}
{Kosugi}, T., et al.
 2007, \solphys, 243, 3

\bibitem[{{Kurokawa} {et~al.}(1987){Kurokawa}, {Hanaoka}, {Shibata}, \&
  {Uchida}}]{Kurokawa.untwist-filamt1987SoPh..108..251K}
{Kurokawa}, H., {Hanaoka}, Y., {Shibata}, K., \& {Uchida}, Y. 1987, \solphys,
  108, 251

\bibitem[{{Moreno-Insertis} {et~al.}(2008){Moreno-Insertis}, {Galsgaard}, \&
  {Ugarte-Urra}}]{Moreno-Insertis.EISjet2008ApJ...673L.211M}
{Moreno-Insertis}, F., {Galsgaard}, K., \& {Ugarte-Urra}, I. 2008, \apjl, 673,
  L211

\bibitem[{{Newton}(1934)}]{NewtonH.surge-discovery.1934MNRAS..94..472N}
{Newton}, H.~W. 1934, \mnras, 94, 472

\bibitem[{{Nishizuka} {et~al.}(2008){Nishizuka}, {Shimizu}, {Nakamura},
  {Otsuji}, {Okamoto}, {Katsukawa}, \&
  {Shibata}}]{Nishizuka.giantCaHjet.2008ApJ...683L..83N}
{Nishizuka}, N., {Shimizu}, M., {Nakamura}, T., {Otsuji}, K., {Okamoto}, T.~J.,
  {Katsukawa}, Y., \& {Shibata}, K. 2008, \apjl, 683, L83

\bibitem[{{Nitta} {et~al.}(2008){Nitta}, {Mason}, {Wiedenbeck}, {Cohen},
  {Krucker}, {Hannah}, {Shimojo}, \&
  {Shibata}}]{NittaN.jetHe3.2008ApJ...675L.125N}
{Nitta}, N.~V., {Mason}, G.~M., {Wiedenbeck}, M.~E., {Cohen}, C.~M.~S.,
  {Krucker}, S., {Hannah}, I.~G., {Shimojo}, M., \& {Shibata}, K. 2008, \apjl,
  675, L125

\bibitem[{{Ofman} \& {Wang}(2008)}]{Ofman.Wang.SOT-wave.2008A&A...482L...9O}
{Ofman}, L. \& {Wang}, T.~J. 2008, \aap, 482, L9

\bibitem[{{Okamoto} {et~al.}(2007){Okamoto}, {Tsuneta}, {Berger}, {Ichimoto},
  {Katsukawa}, {Lites}, {Nagata}, {Shibata}, {Shimizu}, {Shine}, {Suematsu},
  {Tarbell}, \& {Title}}]{Okamoto.Alfven-wave.prominence.2007Sci...318.1577O}
{Okamoto}, T.~J., et al.
 2007, Science, 318, 1577

\bibitem[{{Pariat} {et~al.}(2009){Pariat}, {Antiochos}, \&
  {DeVore}}]{PariatE.twist-jet-MHD.2009ApJ...691...61P}
{Pariat}, E., {Antiochos}, S.~K., \& {DeVore}, C.~R. 2009, \apj, 691, 61

\bibitem[{{Patsourakos} {et~al.}(2008){Patsourakos}, {Pariat}, {Vourlidas},
  {Antiochos}, \& {Wuelser}}]{Patsourakos.EUVI-jet2008ApJ...680L..73P}
{Patsourakos}, S., {Pariat}, E., {Vourlidas}, A., {Antiochos}, S.~K., \&
  {Wuelser}, J.~P. 2008, \apjl, 680, L73

\bibitem[{{Roy}(1973{\natexlab{a}})}]{Roy.surge.dynamics.1973SoPh...32..139R}
{Roy}, J.~R. 1973{\natexlab{a}}, \solphys, 32, 139

\bibitem[{{Roy}(1973{\natexlab{b}})}]{Roy.surge.magnetic.1973SoPh...28...95R}
{Roy}, J.~R. 1973{\natexlab{b}}, \solphys, 28, 95

\bibitem[{{Rust}(1968)}]{RustD.surge-reconn.1968IAUS...35...77R}
{Rust}, D.~M. 1968, in IAU Symposium, Vol.~35, Structure and Development of
  Solar Active Regions, ed. K.~O. {Kiepenheuer}, 77

\bibitem[{{Savcheva} {et~al.}(2007){Savcheva}, {Cirtain}, {Deluca},
  {Lundquist}, {Golub}, {Weber}, {Shimojo}, {Shibasaki}, {Sakao}, {Narukage},
  {Tsuneta}, \& {Kano}}]{Savcheva.XRT.jet.stat2007PASJ...59S.771S}
{Savcheva}, A., et al.
 2007, \pasj, 59, 771

\bibitem[{{Schmieder} {et~al.}(1988){Schmieder}, {Mein}, {Simnett}, \&
  {Tandberg-Hanssen}}]{SchmiederB.cool-surge.hot-jet1988A&A...201..327S}
{Schmieder}, B., {Mein}, P., {Simnett}, G.~M., \& {Tandberg-Hanssen}, E. 1988,
  \aap, 201, 327

\bibitem[{{Shibata} {et~al.}(1992){Shibata}, {Ishido}, {Acton}, {Strong},
  {Hirayama}, {Uchida}, {McAllister}, {Matsumoto}, {Tsuneta}, {Shimizu},
  {Hara}, {Sakurai}, {Ichimoto}, {Nishino}, \&
  {Ogawara}}]{ShibataK.1st-SXT-jet.1992PASJ...44L.173S}
{Shibata}, K., et al.
 1992, \pasj, 44, L173

\bibitem[{{Shibata} {et~al.}(2007){Shibata}, {Nakamura}, {Matsumoto}, {Otsuji},
  {Okamoto}, {Nishizuka}, {Kawate}, {Watanabe}, {Nagata}, {UeNo}, {Kitai},
  {Nozawa}, {Tsuneta}, {Suematsu}, {Ichimoto}, {Shimizu}, {Katsukawa},
  {Tarbell}, {Berger}, {Lites}, {Shine}, \&
  {Title}}]{Shibata.CaHjet.2007Sci...318.1591S}
{Shibata}, K., et al.
 2007, Science, 318, 1591

\bibitem[{{Shibata} \&
  {Uchida}(1985)}]{Shibata.Uchida.helic-jetMHD.1985PASJ...37...31S}
{Shibata}, K. \& {Uchida}, Y. 1985, \pasj, 37, 31

\bibitem[{{Shibata} \&
  {Uchida}(1986)}]{Shibata.Uchida.helic-jet.1986SoPh..103..299S}
---. 1986, \solphys, 103, 299

\bibitem[{{Shimojo} {et~al.}(1996){Shimojo}, {Hashimoto}, {Shibata},
  {Hirayama}, {Hudson}, \& {Acton}}]{ShimojoM.jet-stat.1996PASJ...48..123S}
{Shimojo}, M., {Hashimoto}, S., {Shibata}, K., {Hirayama}, T., {Hudson}, H.~S.,
  \& {Acton}, L.~W. 1996, \pasj, 48, 123

\bibitem[{{Shimojo} {et~al.}(2007){Shimojo}, {Narukage}, {Kano}, {Sakao},
  {Tsuneta}, {Shibasaki}, {Cirtain}, {Lundquist}, {Reeves}, \&
  {Savcheva}}]{ShimojoM.XRT-fine-jet.2007PASJ...59S.745S}
{Shimojo}, M., et al.
 2007, \pasj, 59, 745

\bibitem[{{Shimojo} \& {Shibata}(2000)}]{ShimojoM.jet-phys.2000ApJ...542.1100S}
{Shimojo}, M. \& {Shibata}, K. 2000, \apj, 542, 1100

\bibitem[{{Strong} {et~al.}(1992){Strong}, {Harvey}, {Hirayama}, {Nitta},
  {Shimizu}, \& {Tsuneta}}]{StrongK.SXT-jet1992PASJ...44L.161S}
{Strong}, K.~T., {Harvey}, K., {Hirayama}, T., {Nitta}, N., {Shimizu}, T., \&
  {Tsuneta}, S. 1992, \pasj, 44, L161

\bibitem[{{Tarbell} {et~al.}(1999){Tarbell}, {Ryutova}, {Covington}, \&
  {Fludra}}]{Tarbell.Ryutova.jet.1999ApJ...514L..47T}
{Tarbell}, T., {Ryutova}, M., {Covington}, J., \& {Fludra}, A. 1999, \apjl,
  514, L47

\bibitem[{{Tsuneta} {et~al.}(2008){Tsuneta}, {Ichimoto}, {Katsukawa}, {Nagata},
  {Otsubo}, {Shimizu}, {Suematsu}, {Nakagiri}, {Noguchi}, {Tarbell}, {Title},
  {Shine}, {Rosenberg}, {Hoffmann}, {Jurcevich}, {Kushner}, {Levay}, {Lites},
  {Elmore}, {Matsushita}, {Kawaguchi}, {Saito}, {Mikami}, {Hill}, \&
  {Owens}}]{Tsuneta.SOT.2008SoPh..249..167T}
{Tsuneta}, S., et al.
  2008, \solphys, 249, 167

\bibitem[{{Xu} {et~al.}(1984){Xu}, {Yin}, \&
  {Ding}}]{XuAA.surge-rotate.1984AcASn..25..119X}
{Xu}, A.-A., {Yin}, S.-Y., \& {Ding}, J.-P. 1984, Acta Astronomica Sinica, 25,
  119

\bibitem[{{Yokoyama} \&
  {Shibata}(1995)}]{YokoyamShibata.jetModel1995Natur.375...42Y}
{Yokoyama}, T. \& {Shibata}, K. 1995, \nat, 375, 42

\bibitem[{{Yokoyama} \&
  {Shibata}(1996)}]{YokoyamShibata.jetModel1996PASJ...48..353Y}
---. 1996, \pasj, 48, 353

\end{thebibliography}

}


\end{document}